\documentclass[a4paper]{article}

\usepackage[english]{babel}
\usepackage[utf8x]{inputenc}
\usepackage{hyperref}
\usepackage{amsmath}
\usepackage{amsfonts}
\usepackage{amssymb}
\usepackage{graphicx}
\usepackage{cite}
\usepackage[colorinlistoftodos]{todonotes}

\usepackage[parfill]{parskip} % <-- avoid the  at each paragraph
\usepackage{xspace}

\usepackage{tikz}
\usetikzlibrary{arrows,decorations.pathmorphing,backgrounds,positioning,fit,calc}

\title{Relatedness of the Incidence Decay with Exponential Adjustment (IDEA) Model, ``Farr's Law'' and Compartmental Difference Equation SIR Models}

\author {Mauricio Santillana \and  Ashleigh Tuite \and Tahmina Nasserie \and
Paul Fine \and David Champredon \and Leonid Chindelevitch \and Jonathan Dushoff \and David Fisman}

\date{\today}

\newcommand{\Ro}{\ensuremath{\mathcal{R}_0}\xspace}
\newcommand{\RoIDEA}{\ensuremath{\mathcal{R}_{0,IDEA}}\xspace}
\newcommand{\RoSIR}{\ensuremath{\mathcal{R}_{0,SIR}}\xspace}

\newcommand{\Reff}{\ensuremath{\mathcal{R}_e(t)}\xspace}

\newcommand{\DC}[1]{\textbf{\textcolor{orange}{[DC: #1]}}}

\begin{document}
\maketitle

\begin{abstract}

Mathematical models are often regarded as recent innovations in the description and analysis of infectious disease outbreaks and epidemics, but simple models have been in use for projection of epidemic trajectories for more than a century.  We recently described a single equation model (the incidence decay with exponential adjustment, or IDEA, model) that can be used for short term forecasting.  In the mid-19th century, Dr. William Farr developed a single equation approach (“Farr’s law”) for epidemic forecasting.  We show here that the two models are in fact identical, and can be expressed in terms of one another, and also in terms of a susceptible-infectious-removed (SIR) compartmental model with improving control.  This demonstrates that the concept of the reproduction number $(\Ro)$ is implicit to Farr’s (pre-microbial era) work, and also suggests that control of epidemics, whether via behavior change or intervention, is as integral to the natural history of epidemics as is the dynamics of disease transmission.
% * <david.champredon@gmail.com> 2016-01-08T07:04:03.223Z:
%
% ^.
% * <david.champredon@gmail.com> 2016-01-08T07:03:53.297Z:
%
% > of epidemic trajectories
%
% ^.
\end{abstract}

\section{Introduction}

The control of communicable diseases is an endeavor that has witnessed remarkable successes over the past century; diseases that previously caused large scale mortality have been eradicated \cite{hinman,roeder}, locally eliminated \cite{papania}, or have been markedly reduced in incidence globally as a result of vaccination, antimicrobial therapy, water and sewage treatment, and advances in food safety \cite{armstrong, murray, liu}.  Nonetheless, the threat of communicable diseases persists; emerging infectious diseases continue to be identified, often in association with changes in human and animal mobility, agricultural practices, environmental degradation, and misuse of antimicrobial therapy\cite{kuehn, keesing, jones}. Recent outbreaks or epidemics associated with MERS coronavirus \cite{mers}, influenza A(H7N9) \cite{cowling}, and the West African emergence of the Zaire strain of Ebola virus \cite{ebola}, have challenged epidemiologists as the natural history, modes of transmission, and/or means of control of these diseases have not been well understood during initial periods of emergence.\\

When novel infectious diseases emerge or familiar diseases resurge, mathematical models can serve as useful tools for the synthesis of available data, management of uncertainty, and projection of likely epidemic trajectories \cite{paninform}.  
While it may be challenging to parameterize detailed mechanistic mathematical models when there is little information on mechanisms of transmission, baseline immunity in a community, the nature of the infecting pathogen, et cetera, a number of descriptive approaches exist which may permit fitting, and forecasting, of an epidemic curve.  One single equation approach that has been applied to emerging infections is the Richards model, which treats cumulative infections as a logistic growth process \cite{hsieh}.  However, the concept of modeling an epidemic curve as a simple function, without reference to mechanisms of transmission, is in fact much older, and may originate in the work of the English polymath Dr. William Farr (1807-1883), who rose from humble beginnings to become a physician, mathematician, hygienist and protege of Lancet founder Dr. Thomas Wakley \cite{fine,brownlee1, greenwood1}.  Dr. Farr spent almost 40 years at the General Register Office of the United Kingdom, and the esteem in which he was held is apparent in the "letters" he published annually as appendices to the reports of the Registrar General, in which he supplemented  the dry statistical reports with thoughtful and creative musings on topics as wide-ranging as the relationships between occupation and disease, suicide and mortality in the mentally ill, population density and mortality, and as above the "laws" governing epidemics \cite{farr_letter_sp}.\\

\begin{figure}
\centering
\includegraphics[width=1\textwidth]{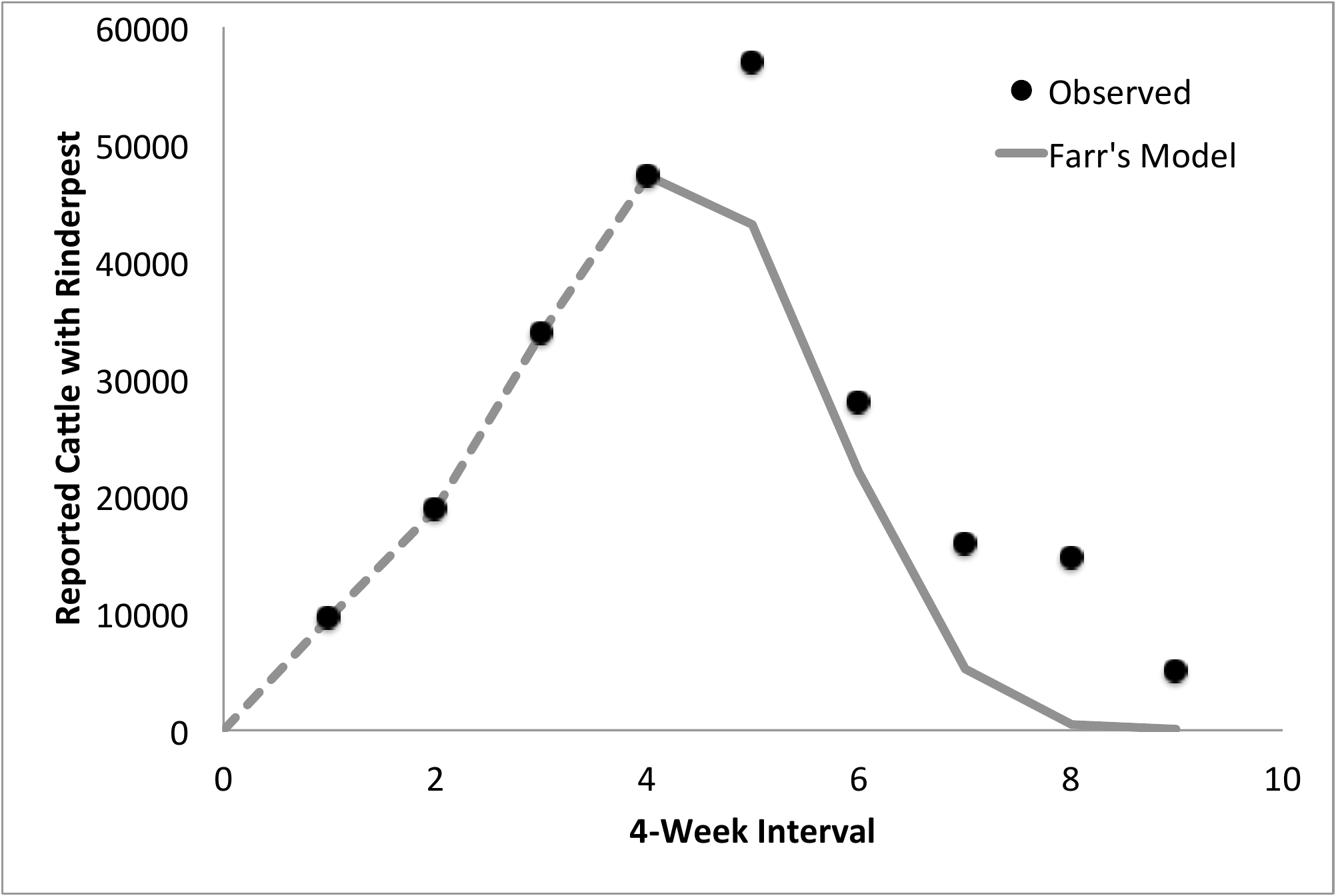}
\caption{A recreation of Farr's possible approach to the English rinderpest epidemic of 1865-1866, based on data presented in \cite{brownlee1}.  Black dots = empirical data, whereas gray curve represent Farr's "model projections".  Dashed component of curve represents fitted component of "model'', while solid gray curve represents Farr's projection.  Initial data point is November 4, 1865, with subsequent points presented at 4 week intervals to June 16, 1866.}
\label{fig:rinderpest}
\end{figure}

Using what was subsequently dubbed "Farr's law'' \cite{greenwood1,langmuir}, Farr used the ratio of case counts in successive time periods to successfully forecast the size and duration of epidemics of smallpox and rinderpest \cite{brownlee1, greenwood1} (Figure \ref{fig:rinderpest}).  Farr demonstrated in a "letter" of 1837 that the decline of a national smallpox epidemic was exponential rather than linear \cite{farr_letter_sp}.  He used related methods over 25 years later, to refute the assertion by a British parliamentarian that a rinderpest epidemic in British cattle would continue to grow exponentially over time\cite{brownlee1,greenwood1}.  Again, Farr demonstrated that although case numbers were increasing markedly, the rate of increase was, in fact, decelerating; he asserted that the epidemic would be large but would end in the coming months, a projection which proved to be correct.\\  

Farr was famously vague about the mathematical underpinnings of his projections, though key to his observation was the observation that case counts would decrease by a constant log-quantity following initial exponential epidemic growth \cite{brownlee1}.  It fell to other contemporaries \cite{evans} and later epidemiologists (most notably Dr. John Brownlee) to formalize "Farr's law" \cite{fine,brownlee2,serfling}.  (It should be noted that the term "Farr's law'' is ambiguous.  Farr himself referred to a "law" in his letter on rinderpest \cite{brownlee1}, but the term has also been used by others to describe Farr's observations on the relation between population density and death \cite{brownlee_mort}, and to his description of the relationship between cholera mortality and altitude \cite{lilienfeld}.  In his elaboration of the "law'', Brownlee referred to it as "Farr's theory of epidemics''\cite{brownlee1}).\\

We recently proposed a descriptive approach to the initial estimation of the basic reproduction number $(\Ro)$ of an emerging or re-emerging pathogen, which also provides information on the rate at which the process is being controlled, as well as reasonable short-term projections of incidence.  
This two-parameter model, which we have referred to as the ``Incidence Decay with Exponential Adjustment'' (IDEA) model, offers advantages of simplicity, explicit linkage to theory of epidemic growth, and also acknowledges the fact that epidemics and outbreaks do not peak and end simply due to depletion of susceptibles, but because of a complex constellation of public health actions and behavioral changes that may modify the course of an epidemic and reduce the effective reproductive number \Reff of an outbreak \cite{hauck}.  
One of us (PF) had previously written about Farr's rule and its importance in the development of epidemic theory\cite{fine}, and noted the conceptual similarity between IDEA and Farr's rule.  Upon exploration of these two approaches we realized that they are, notwithstanding having been formulated some 160 years apart, and being based on very different theoretical constructs, identical. Here we demonstrate the equivalence of these methods for epidemic modeling and forecasting, and explore the potential advantages of using these methodologies to complement other approaches. We explore numerical examples derived from current emerging infectious diseases, including the West African Ebola virus outbreak. We show that these methodologies work fairly well in a number of cases, and  discuss what this fact implies about difficult-to-measure components of transmission and control parameters.

\section{Methods}

\subsection{Farr's Rule}

The empirical relationship between observed cases of infected individuals, in sequential time intervals, during an epidemic outbreak according to \textbf{Farr's rule} is given by
\begin{equation}
\frac{\dfrac{I(t+3)}{I(t+2)}}{\dfrac{I(t+1)}{I(t)}}=K
\label{Farr}
\end{equation}
where $I(t)$ represents the number of observed newly infected individuals at time $t$, and $K$ is a constant. For values of $K<1$, the rate of change in the observed cases (acceleration) decreases as time evolves, and the family of curves $I(t)$ that satisfy equation (\ref{Farr}) correspond to familiar bell-shaped epidemiological curves.
\paragraph{}The rule was investigated and elaborated upon by the epidemiologist John Brownlee who noted that $I(t)$ under this rule would correspond to the function \cite{fine}\\ 
\begin{equation} 
\exp(-At^2+Bt+C)
\label{Brownlee}
\end{equation} 
where $A$, $B$, and $C$ are constants. Of note, Brownlee's formulation identifies a process where cases increase as a first order process, but decrease as a second order process, as is the case with the IDEA model. 

\subsection{IDEA Model}

In its basic form the IDEA model holds
\begin{equation}
I(t)=\left(\frac{\Ro}{(1+d)^t} \right)^t
\label{IDEA}
\end{equation}
where \(I(t)\) is the incident infections in generation \(t\) of an outbreak (thus \(t=\left\lbrace 1,2,3, ...\right\rbrace\in\mathbb{N}\)).  This is in contradistinction to equation (1) above where $t$ is an arbitrary, but constant, time interval (not necessarily equal to a generation interval).  
The parameter \Ro is the basic reproduction number as usually defined; that is, the number of secondary cases created by a primary case in a totally susceptible population and in the absence of intervention.  The parameter $d$ which we have referred to as a ``control parameter'' defines the rate at which transmission declines over the course of an epidemic.  
The empirical underpinnings of $d$ are not yet well defined, but based on current understanding of disease dynamics could represent public health interventions, population adaptation or behavior change, improved availability of personal protective items or effect of drugs to treat infection, or reductions in population susceptibility as a result of infection or vaccination.  
As noted above, by fitting the model to data we have previously obtained reasonable estimates of \Ro early in the course of epidemics, and have also been able to produce plausible near-term projections of future case counts.

\subsection{Difference Equation Susceptible-Infectious-Removed Model}

In an earlier publication we commented on the almost identical projections generated by IDEA and a compartmental difference equation (Susceptible-Infectious-Removed, or SIR) \cite{hauck}, when \Ro is small and when there is exponential reduction in risk over time.  Indeed, the effect we identified can be generalized to any situation where depletion of susceptibles due to infection is small relative to the total population size, and not only when \Ro is small. We used a ``damped'' version of the standard SIR model whose formulation in generation interval time scale is given by:

\begin{eqnarray}
\label{eq:SIR}
S_{t+1} & = & S_t - \Reff I_t \\
I_{t+1} & = & \Reff I_t
\end{eqnarray}

with $S_t$ the number of susceptible individuals at time $t$ and \Reff the effective reproductive number at time $t$ defined by:

\begin{equation}
\label{eq:ReffDamped}
\Reff=\Ro\frac{S_t}{N}\rho^{t}
\end{equation}

The ``dampening'' parameter $\rho$ represents the relative risk of infection in each generation of the epidemic, compared to risk seen in the last generation (i.e., if there were no improvement in control in a given generation compared to the last, most recent generation).  If an outbreak is small relative to the size of the total population (as would be true if \Ro were modest and control achieved relatively quickly)  $S(t)/N$ will be approximately 1 throughout the outbreak and the expression can be rewritten as:
\begin{equation}
I(t)=I(0) \prod_{s<t}\left( \rho^s \Ro \right)
\end{equation}
 meaning that all reduction in \Reff is due to control rather than depletion of susceptibles.  \Ro is a constant, and the sum of exponents of $\rho$ is simply $t(t+1)/2$.  We can assume (as we do with IDEA) that the outbreak began with the introduction of a single case such that $I(0)$ = 1.  Now:

\begin{equation}
I(t)=\rho^{t(t+1)/2}\Ro^{t}
\end{equation}

\subsection{Numerical Examples using Recent Outbreak Data}

We have used IDEA to explore the nature of epidemic growth during the recent West African Ebola epidemic\cite{plos1,plos2}, the 2014 MERS coronavirus outbreak in Saudi Arabia\cite{plos3}, and more recently Chikungunya virus invasion in the Western hemisphere \cite{nasserie}.  As publicly available data have taken the form of cumulative incidence curves (with absent dates of onset) we fit IDEA to cumulative curves, but it is possible to estimate "pseudo-incidence'' by taking the interval to interval difference cumulative incidence over time.  

We fit IDEA to the incidence time series and calculated Farr's $K$ for sequential generation tetrads, and converted $K$ values to the $d$ parameter in IDEA using the relation $K =\frac{1}{(1+d)^4}$ described below.  
Of necessity, we excluded $K$ estimates derived from sequences of intervals containing negative incidence estimates (i.e., those where reported cumulative incidence declined). Data sources used for these analyses are available at \\
\href{url}{http://figshare.com/authors/Tahmina\_Nasserie/686527}(Chikungunya) and \\
\href{url}{https://github.com/cmrivers/ebola} (Ebola).

\section{Results}

\subsection{Equivalence of IDEA and Farr's Rule}

An incidence curve described by IDEA naturally satisfies Farr's rule. Indeed, substituting (\ref{IDEA})  into (\ref{Farr}) gives
\begin{equation}
\label{eq:IDEA_FARR}
K=\frac{1}{(1+d)^4}
\end{equation}
It is interesting to note that the value of \Ro in the IDEA model is irrelevant in the proof (see Appendix \ref{appendix:IDEA_FARR}).

Moreover, by expressing the IDEA model as
 \begin{equation}
I(t)=\left(\frac{\Ro}{(1+d)^t} \right)^t=\exp \biggl(t \log \frac{\Ro}{(1+d)^t}  \biggr) =\exp \bigl(-t^2 \log(1+d) + t\log {\Ro} \bigr) \label{IDEA2}
\end{equation}
 
we see that we recover John Brownlee's Gaussian curve (\ref{Brownlee}) with $A=\log(1+d)$, $B=\log\Ro$ and $C=0$.

\subsection {\textit{K} as an Odds Ratio}

As $K$ represents a ratio of ratios, it can be conceptualized as equivalent in form to an odds ratio. Effectively, $\frac{I_t}{I_{t+1}}$ is the odds of a case occurring in an initial as opposed to a subsequent generation while $K$ then becomes interpretable as an odds ratio, though it is unclear whether this odds ratio has an intuitive meaning.  Nonetheless, this form is important as it suggests that the asymptotic variance of $\log(K)$ can be estimated as $(\frac{1}{I_t} + \frac{1}{I_{t+1}} + \frac{1}{I_{t+2}} + \frac{1}{I_{t+3}})$ \cite{rothman}. Estimation of variance of $\log(K)$ makes it possible to estimate confidence limits for $K$, and thus for $d$.  Furthermore, given that several $K$ estimates can be generated from a given epidemic curve, estimation of variance should permit the use of Mantel-Haenszel methods \cite{mantelhaenszel} to generate summary estimates of $K$ over the course of an epidemic, or to use meta-regressive methods to evaluate $K$ for trends \cite{metareg}.  A potential pitfall here is the non-independence of serial estimates of $K$ due to overlap in incidence values used in adjacent estimates of $K$.

\subsection{Relation of SIR Model to Farr's \textit{K} and IDEA}

When the depletion of susceptible is negligible compared to the total population size (that is, a small outbreak), we can actually express IDEA's basic reproductive number \RoIDEA  and its control factor $d$ as a function of the basic reproductive number \RoSIR and the control factor $\rho$ of the damped SIR model described in (\ref{eq:SIR}-\ref{eq:ReffDamped}). The relationship between the parameters of these two models is (details in Appendix \ref{appendix:IDEA_SIR}) 

\begin{equation}
\label{eq:R0_IDEA_SIR}
 \RoIDEA \simeq \frac{\RoSIR}{\sqrt{\rho}} 
\end{equation}
\begin{equation}
\label{eq:d_rho}
d \simeq \frac{1}{\sqrt{\rho}} -1
\end{equation}

Substituting (\ref{eq:d_rho}) in equation (\ref{eq:IDEA_FARR}) we can link Farr's rule with the damped SIR model:
\begin{equation}
\label{eq:FARR_SIR}
K \simeq \rho^2
\end{equation}

\subsection{Numerical Simulations}

In this section, we aim to test numerically the validity of approximations (\ref{eq:R0_IDEA_SIR}) and (\ref{eq:d_rho}). In particular, given a damped SIR model, we explore the parameter space \RoSIR and $\rho$ for which these approximations hold. Note that the link between IDEA and Farr's rule given by equation (\ref{eq:IDEA_FARR}) is not an approximation, but a genuine equality (subject to the time step used in Farr's rule being equivalent to one generation), so there is no need to test it numerically. To measure the performance of the approximation, we consider the distance between the simulated incidence time series. Let $N$ be the number of generations simulated and $I_{SIR}(k)$ (resp. $I_{IDEA}(k)$) the incidence from the SIR (resp IDEA) model at the $k^{th}$ generation, we define their distance by

\begin{equation}
\delta = \sqrt{\sum_{k=1}^N \left( I_{SIR}(k)-I_{IDEA}(k)  \right)^2 }
\end{equation}

The tests are implemented in the software \textsf{R} \cite{citeR} and code is available in electronic supplement files. Figure \ref{fig:paramSpace} shows the values of the distance $\delta$ for different values of \RoSIR and $\rho$. We see that for a combination of \RoSIR and $\rho$ such that the depletion of susceptible is not too large, the approximation is very good. But when the values of \RoSIR and $\rho$ generate a depletion of susceptible individuals that is no longer negligible (white area in Figure  \ref{fig:paramSpace}), then the incidence curve from the IDEA model diverges from that generated by the damped SIR model.

\subsection{Application: Ebola and Chikungunya}

Considering real epidemic data from Ebola and Chinkungunya, it can be seen that the interval to interval variability in $K$ was substantial, likely reflecting variability in reporting (Figure 5 and Figure 6).  Simple arithmetic means of $K$ over time were also unstable due to skewing by values substantially greater than 1.  However, when we estimated the geometric mean of $K$ over time we found that the resultant $d$ estimate approximated that derived through fitting IDEA.  

Furthermore, we noted that in the Chikungunya time series there was a large perturbation in best-fit values of $d$ occurring in October 2014, corresponding with an apparent multi-wave epidemic.  We have previously noted that this abrupt change the generation-to-generation best fit value of $d$ corresponds with the occurrence of multiwave epidemics when IDEA is fit to simulated data \cite{hauck}; using Farr's approach, the onset of a possible new Chikungunya wave seems to correspond with an abrupt increase in $K$ to a value far greater than one  (Figure 7).  The utility of large values of $K$ as a signal of an incipient epidemic wave warrants further investigation.

\section{Discussion}

Although the real-time application of mathematical modeling to understanding and control of outbreaks is often perceived as representing a recent development in infectious disease epidemiology\cite{heesterbeek}, disease modeling has deeper historical roots, including work by Bernoulli on smallpox in the 18th century\cite{greenwood2}; work by Ross on malaria transmission\cite{ross}, and as mentioned above, Farr’s work on the growth and cessation of epidemics  \cite{brownlee1, brownlee2, fine}.  We had published a simple, phenomenological approach to the description and projection of outbreaks and epidemics \cite{hauck} which we had initially regarded as a novel formulation rooted in the concept of the basic reproduction number \Ro.  

In that work, we demonstrated concordance with projections derived using a 3-compartment difference equation model (damped SIR model).  We have subsequently realized, and demonstrate above, that our approach simply represented a restatement of Farr’s work, albeit in a manner that is tied to the concept of \Ro.  According to Brownlee, Farr promised to describe the derivation of his model in greater detail in future reports, but never did so \cite{brownlee1}, and much of the mathematical elaboration of Farr’s work was in fact done by Brownlee after Farr’s death\cite{brownlee2}.  Nonetheless, Brownlee notes that to Farr, the predictive accuracy of his approach reflected three characteristics of epidemics, according to Farr’s (pre-microbial) understanding: (i) diminution in the number of susceptibles over time due to recovery from infection (“immunity”, though to use this term in application to Farr is an anachronism); (ii) diminished population density due to death from infection; and (iii) diminishing pathogenicity of the disease with each passing generation of infection as a result of (to quote Farr) "[loss of] part of the force of infection in every body through which they pass...the matter...diminishes in strength at every transmission by innoculation"\cite{brownlee1}.  The first two characteristics are not incompatible with the current understanding of epidemic dynamics, whereas the third is not  (though it does anticipate more modern ideas around evolution of virulence and disease ecology \cite{lipsitch, ewald}).

However, as this model is phenomenological, rather than mechanistic in nature, the putative epidemiological mechanisms underlying model performance are not of immediate importance.  Indeed, while the simplicity of this approach may be regarded as a limitation, the simplicity of the form, and its implicit incorporation of biological, social, medical, and behavioral drivers of control into a single parameter estimated via fitting, may be a strength, especially given that such control factors as behavior change due to fear may be difficult or impossible to measure in real time.

When we have applied IDEA to current day outbreaks and epidemics, we have remained agnostic about the factors that cause second order deceleration of epidemic growth.  Referring to first principles, the components of a reproduction number are duration of infectiousness, contact rate, and probability of transmission conditional on contact, as well as susceptibility in a population \cite{vynnycky}.  We presume that public health interventions, population behavior change (as a result of education or rumours, prudence or fear), and the occurrence of silent infections with immunity could all contribute to deceleration of epidemic spread, even when the effective reproductive number is expected to be greater than 1 due to widespread susceptibility in the population.  Furthermore, we note that Farr's original "time step'' appears to have been arbitrary (reflecting the form of data available to him: weekly for rinderpest, quarterly for smallpox), whereas we have used generations as time steps in our more recent applications of Farr's "law'', and in IDEA and SIR models.  This potential discrepancy between Farr's initial efforts and our more recent efforts warrants further exploration.

\subsection{Summary}

In demonstrating that the IDEA model and Farr’s model are mathematically identical (and can be virtually identical to an SIR model with a small \Ro, abundant susceptibility in the population, and exponential improvement in control) we demonstrate that recognizing the underlying mechanism of epidemic control may be unimportant for generating reasonable forecasts of epidemics with control, or recognizing when their fundamental dynamics have changed.  Our contribution in the current work is to show that Farr’s law, while derived in the pre-microbial era, can be reformulated in terms of the concept of the basic reproductive number, combined with exponential increase in control via whatever mechanism.  We observe, unexpectedly, that Farr’s K can be expressed as a function of the IDEA d parameter alone, independent of \Ro, implying that epidemic trajectory is (and has historically been) more a function of control efforts and changing behavior than of the fundamental characteristics of a given infectious disease.  Whether or not the ratio K can have stand-alone value as a tool to identify unexpected shifts in epidemic trajectory (e.g., the two wave epidemic of Chikungunya referred to in Figure 7 above) will be the subject of future work.\\

\newpage
\section{Appendix}
\subsection{Details on IDEA and Farr's law relationship}
\label{appendix:IDEA_FARR}

\DC{Depending of the journal this manuscript is submitted to, I would either (if slightly technical journal) not include this proof at all or (non-technical journal) do not include case $t=1$}

Assume that the parameters \Ro and $d\geq 0$, in the IDEA model, describe accurately the observed number of infected cases as a function of (serial) time, in an ideal outbreak. Then we can substitute $I(t)$ from equation (\ref{IDEA}) into equation (\ref{Farr}) to see if the sequential number of infected individuals, as predicted by the IDEA model, satisfy Farr's law.\\

For clarity, choose $t=1$, thus equation (\ref{Farr}) becomes
\begin{eqnarray*}
\frac{\dfrac{I(4)}{I(3)}}{\dfrac{I(2)}{I(1)}}=\frac{\dfrac{\left(\frac{\Ro}{(1+d)^4} \right)^4}{\left(\frac{\Ro}{(1+d)^3} \right)^3}}{\dfrac{\left(\frac{\Ro}{(1+d)^2} \right)^2}{\left(\frac{\Ro}{(1+d)} \right)}}&=&
\frac{\left(\dfrac{\Ro^4}{\Ro^3}\right)\left( \dfrac{((1+d)^3)^3}{((1+d)^4)^4}f\right)}{\left(\dfrac{\Ro^2}{\Ro}\right)\left( \dfrac{(1+d)}{((1+d)^2)^2}\right)}\\
&=&
\frac{\Ro \left( \dfrac{(1+d)^9}{(1+d)^{16}}\right)}{\Ro \left( \dfrac{(1+d)}{(1+d)^4}\right)}=
\frac{ (1+d)^{-7}}{(1+d)^{-3}}=\frac{1}{(1+d)^4}
\end{eqnarray*}
By identifying the constant $K=1/(1+d)^4$, the IDEA model satisfies Farr's law for $t=1$. In general, for any sequential (integer) time intervals $t, \;t+1,\; t+2,\; t+3$ one can generalize the above result as follows:
\begin{eqnarray*}
\frac{\dfrac{I({t+3})}{I({t+2})}}{\dfrac{I({t+1})}{I(t)}}=\frac{\dfrac{\left(\frac{\Ro}{(1+d)^{t+3}} \right)^{t+3}}{\left(\frac{\Ro}{(1+d)^{t+2}} \right)^{t+2}}}{\dfrac{\left(\frac{\Ro}{(1+d)^{t+1}} \right)^{t+1}}{\left(\frac{\Ro}{(1+d)^t} \right)^t}}&=&
\frac{\left(\dfrac{\Ro^{t+3}}{\Ro^{t+2}}\right)\left( \dfrac{((1+d)^{t+2})^{t+2}}{((1+d)^{t+3})^{t+3}}\right)}{\left(\dfrac{\Ro^{t+1}}{\Ro^t}\right)\left( \dfrac{((1+d)^t)^t}{((1+d)^{t+1})^{t+1}}\right)}\\
&=&
\frac{\Ro \left( \dfrac{(1+d)^{(t+2)^2}}{(1+d)^{(t+3)^2}}\right)}{\Ro \left( \dfrac{(1+d)^{t^2}}{(1+d)^{(t+1)^2}}\right)}=
\frac{ (1+d)^{(t+2)^2-(t+3)^2}}{(1+d)^{t^2-(t+1)^2}}\\
&=&\frac{ (1+d)^{-2t-5}}{(1+d)^{-2t-1}}=\frac{1}{(1+d)^4}\end{eqnarray*}

\subsection{Details on IDEA and SIR relationship}
\label{appendix:IDEA_SIR}
Incidence or the SIR model (Equation \ref{eq:SIR}) can be written as
\begin{eqnarray*}
I_t^{SIR} & = &   RE(t) I_{t-1} \\
& = & I_0 \prod_{h=0}^{t-1} RE(s)\\
& = & I_0 \prod_{h=0}^{t-1} R_{0,SIR}\,\,S_h/N\rho^h\\
& = & R_{0,SIR}^t\,\, \rho^{t(t-1)/2}\prod_{h=0}^{t-1}S_h/N  \\
\end{eqnarray*}

If the epidemic size is small compared to the size of the whole population, then it can be assumed that 
$$\prod_{h=0}^{t-1}\left(\frac{S_h}{N}\right) \simeq 1$$
In that case, 
\begin{equation}
\label{eq:I_SIR}
I_t^{SIR}  \simeq \RoSIR^t\,\, \rho^{t(t-1)/2} = \frac{(\RoSIR/\sqrt{\rho})^t}{(1/\sqrt{\rho})^{t^2}}
\end{equation}

Incidence in the IDEA framework is simply
\begin{equation}
\label{eq:I_IDEA}
I_t^{IDEA} = \left(\frac{\RoIDEA}{(1+d)^t} \right)^t = \frac{\RoIDEA^t}{(1+d)^{t^2}}
\end{equation}

Finally, both models have the at time $t$ if $I_t^{SIR}=I_t^{IDEA}$. A sufficient condition for that equality to hold is when both numerator and denominator of equations (\ref{eq:I_SIR}) and (\ref{eq:I_IDEA}) are equal, that is 
\begin{equation*}
 \RoIDEA \simeq \frac{\RoSIR}{\sqrt{\rho}} 
\end{equation*}
\begin{equation*}
d \simeq \frac{1}{\sqrt{\rho}} -1
\end{equation*}\\

\newpage
\section{Figures}

%%%%%%%%%%%%%%%%%%%%%%%%%%%%%%%
%%%   DIAGRAM OF THE MODEL USING TIKZ PACKAGE
%%%%%%%%%%%%%%%%%%%%%%%%%%%%%%%

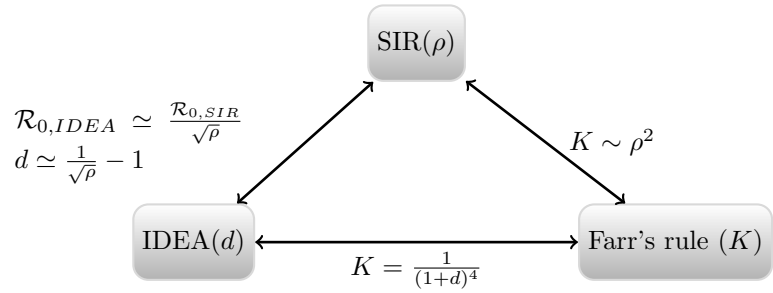
\begin{figure}
\begin{center}
\begin{tikzpicture}
[node distance=20mm,
%
% COMPARTMENT
compartment/.style={
% The shape:
rectangle,
minimum size=10mm,
rounded corners=2mm,
% The rest
thick, 
draw=black!15,
top color=white,bottom color=black!30},
%
% TEXT BOX
mytextbox/.style={
% The shape:
rectangle,
% The rest
text=black!50,
thin, 
draw=white,
top color=white,bottom color=white,
fill=white}
]
% ==== Top Panel ====
\node (SIR)		[compartment] 					{SIR($\rho$)};
\node (empty) 		[mytextbox, below=of SIR] 			{};
\node (IDEA) 		[compartment, left=of empty] 			{IDEA($d$)};
\node (Farr) 	[compartment, right=of empty] 			{Farr's rule ($K$)};

\path	(SIR)		edge[<->,line width=1 pt]	node[right,rotate=0]		{\,\,\,$K \sim \rho^2$}			(Farr)
		(Farr)	edge[<->,line width=1 pt]	node[below,rotate=0]		{$K=\frac{1}{(1+d)^4}$}	(IDEA)
		(IDEA)	edge[<->,line width=1 pt, text width=105]	node[left,rotate=0]		{$\RoIDEA \simeq \frac{\RoSIR}{\sqrt{\rho}}$ $d \simeq \frac{1}{\sqrt{\rho}} -1$ }		(SIR);
\end{tikzpicture}
\caption{Relationship between the three models. Parameter $\rho$ represent the relative risk of infection in each generation for the SIR model. The control parameter $d$ is associated with the IDEA model and $K$ is Farr's ratio.}
\label{fig:extendSEIR}
\end{center}
\end{figure}

\begin{figure}
\centering
\includegraphics[width=1\textwidth]{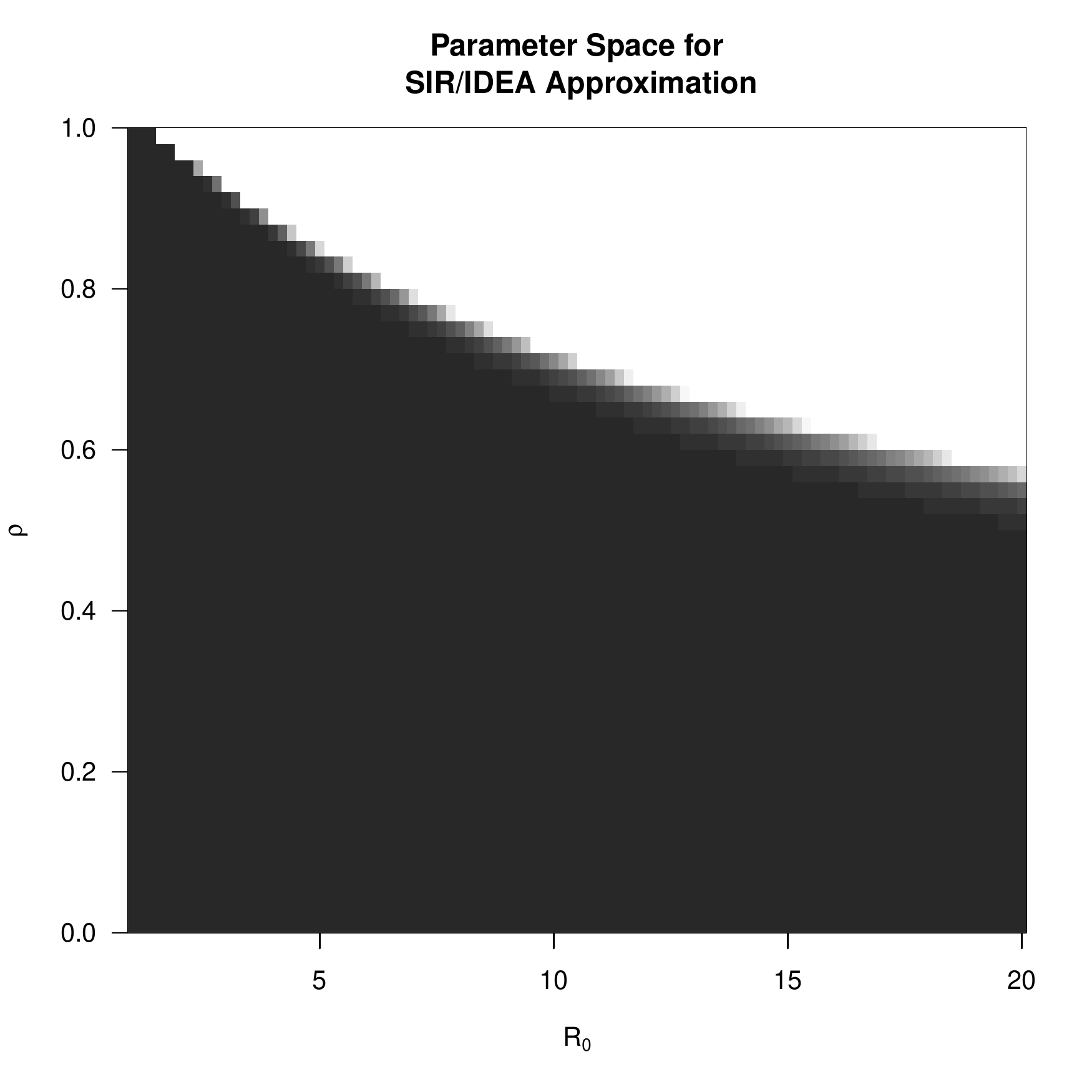}
\caption{A heat map plotting values for \RoSIR and $\rho$ where the damped SIR model can be approximated by a IDEA model using equations (\ref{eq:R0_IDEA_SIR}) and (\ref{eq:d_rho}). Darker areas indicate a good match (measured as the sum of squared differences) between the simulated incidence time series; lighter areas represent combinations of values for which incidence time series for SIR and IDEA diverge.}
\label{fig:paramSpace}
\end{figure}

\newpage
\begin{figure}
\centering
\includegraphics[width=1\textwidth]{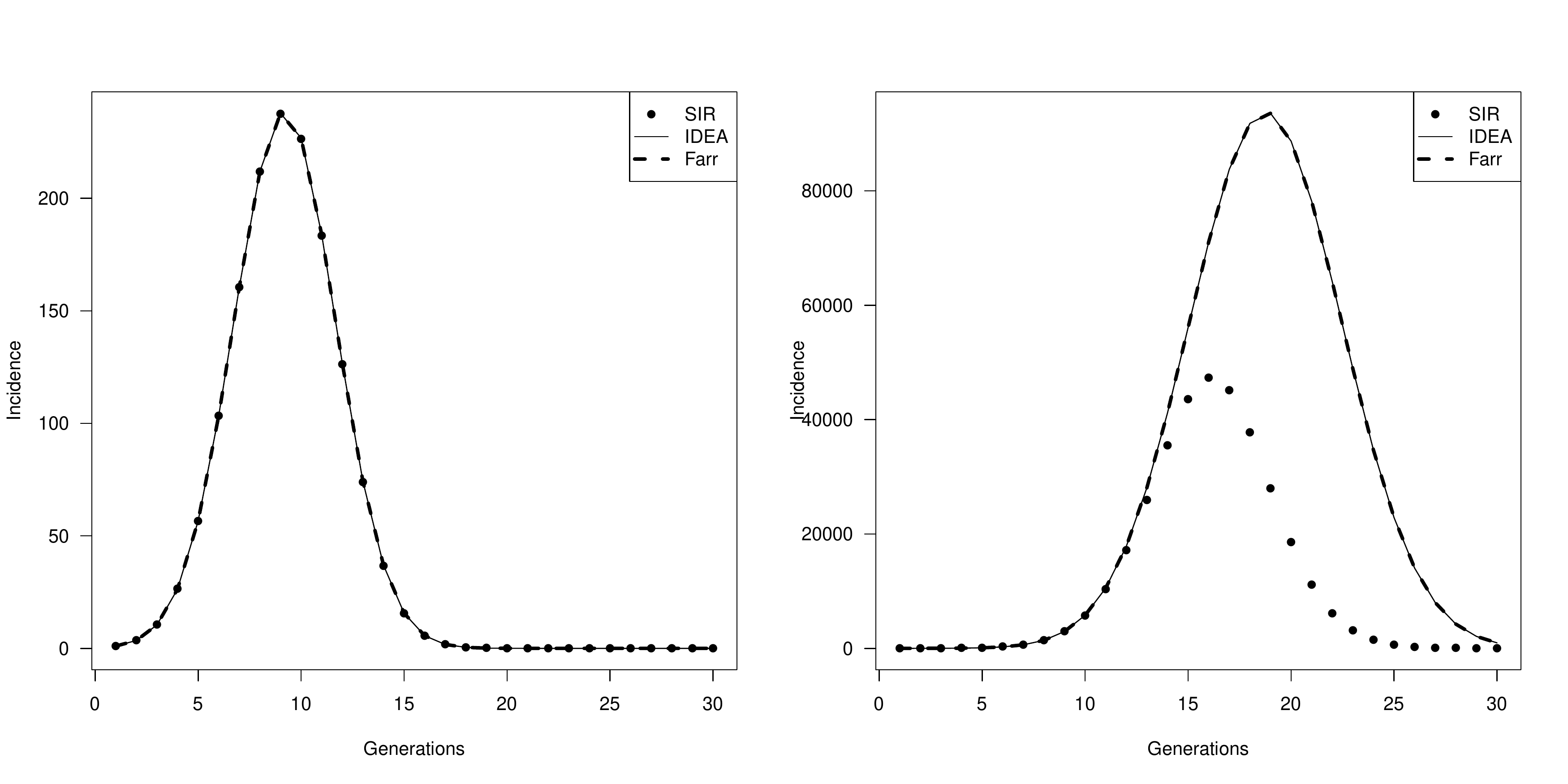}
\caption{\label{fig: Figure 4} Evaluation of IDEA model fit to simulated data derived from the damped SIR model.  Left panel illustrates a scenario where approximation is very good, (\RoSIR=3.5, $\rho=0.85$), corresponding to a combination of values found in the dark area in Figure \ref{fig:paramSpace}. The right sided panel uses a combination of values (high \RoSIR and/or low $\rho$) where susceptible depletion cannot be ignored (i.e., corresponding to the white area in Figure \ref{fig:paramSpace}).  It can be seen that IDEA and the damped SIR models diverge when susceptibles are rapidly depleted.}\end{figure}

\newpage
\begin{figure}
\centering
\includegraphics[width=1\textwidth]{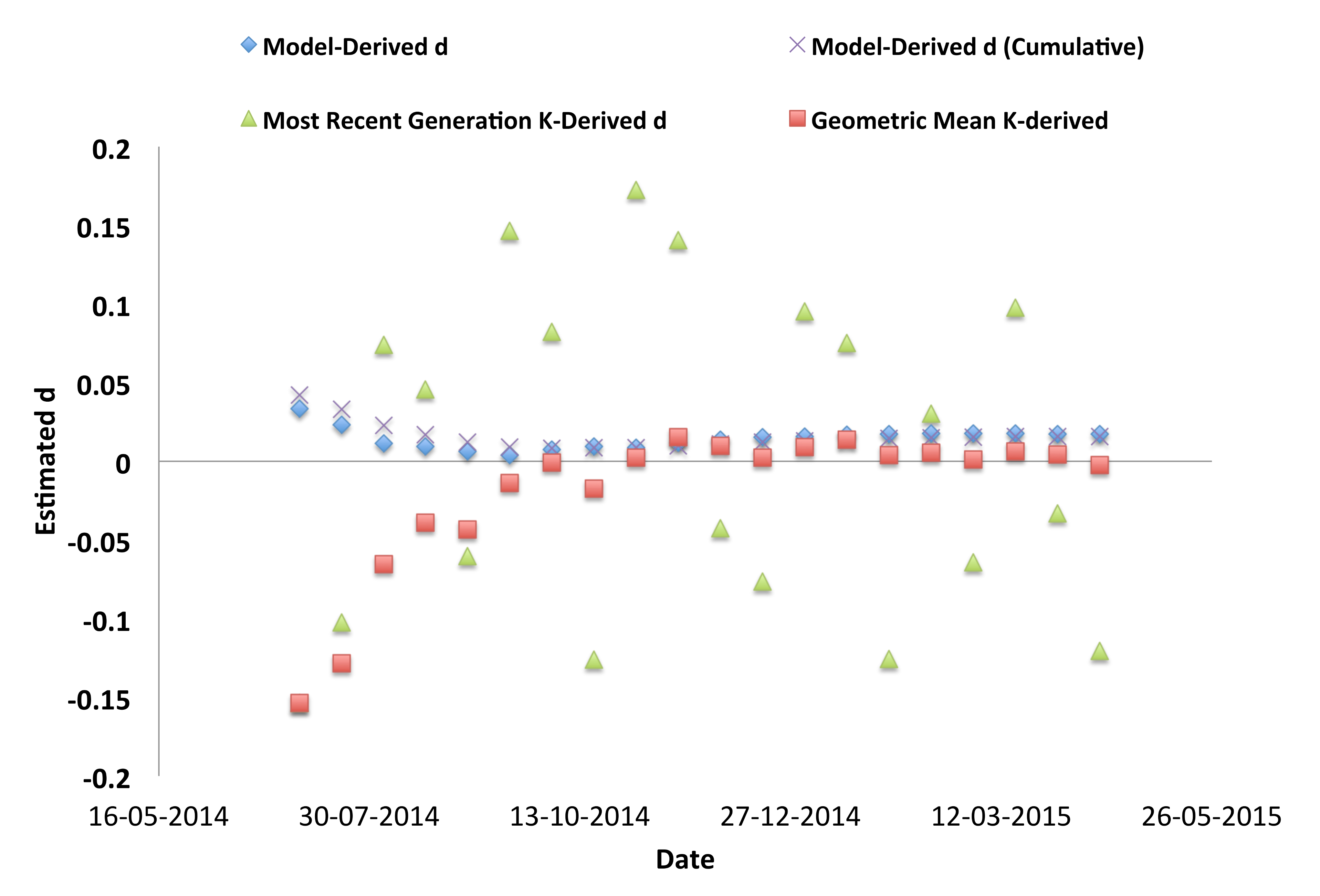}
\caption{\label{fig:Figure 5}The graph plots estimates of IDEA $d$ parameter against time during the recent West African Ebola outbreak.  Approximate date of the last generation incorporated into estimates is plotted on the X-axis; estimated $d$ is plotted on the Y-axis.  $d$ estimates were either derived via IDEA model fitting to ``incident'' cases (blue diamonds) or cumulative incidence (crosses), or derived by estimating Farr's $K$ and transforming resultant estimates using the relation described by equation (\ref{eq:IDEA_FARR}). When $K$ is estimated using 4-generation series (green diamonds), resultant $d$ estimates are volatile and bear little resemblance to $d$ estimates derived through fitting IDEA.  However, estimates of $K$ derived as geometric means of all available $K$ values (red squares) provide a more reasonable approximation of $d$.}
\label{fig:Figure5}
\end{figure}

\newpage
\begin{figure}
\centering
\includegraphics[width=1\textwidth]{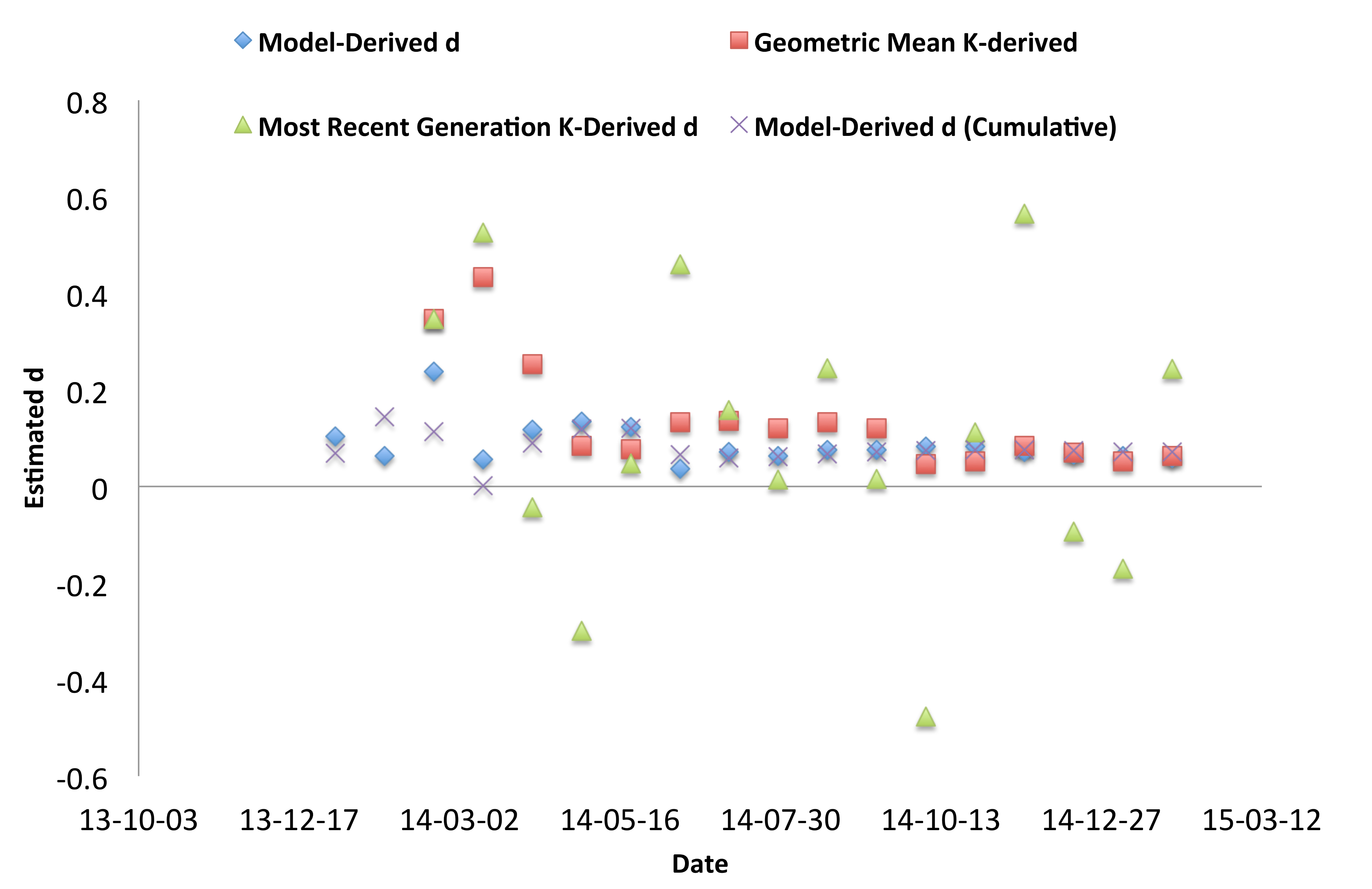}
\caption{\label{fig:Figure 6} As with Figure 5, this figure shows estimates of $d$, derived directly by model fitting or by transforming estimates of Farr's $K$, for the emerging Western Hemisphere Chikungunya epidemic in 2014-2015.  As in Figure 5, $d$ estimates were either derived via IDEA model fitting to ``incident'' cases (blue diamonds) or cumulative incidence (crosses), or derived by estimating Farr's $K$ and transforming resultant estimates.  As in Figure 5, volatile estimates of $K$ were derived using 4-generation series (green diamonds), but estimates of $K$ derived as geometric means of all available $K$ values (red squares) provided a reasonable approximation of $d$.  }
\label{fig:Figure6}
\end{figure}

\newpage
\begin{figure}
\centering
\includegraphics[width=1\textwidth]{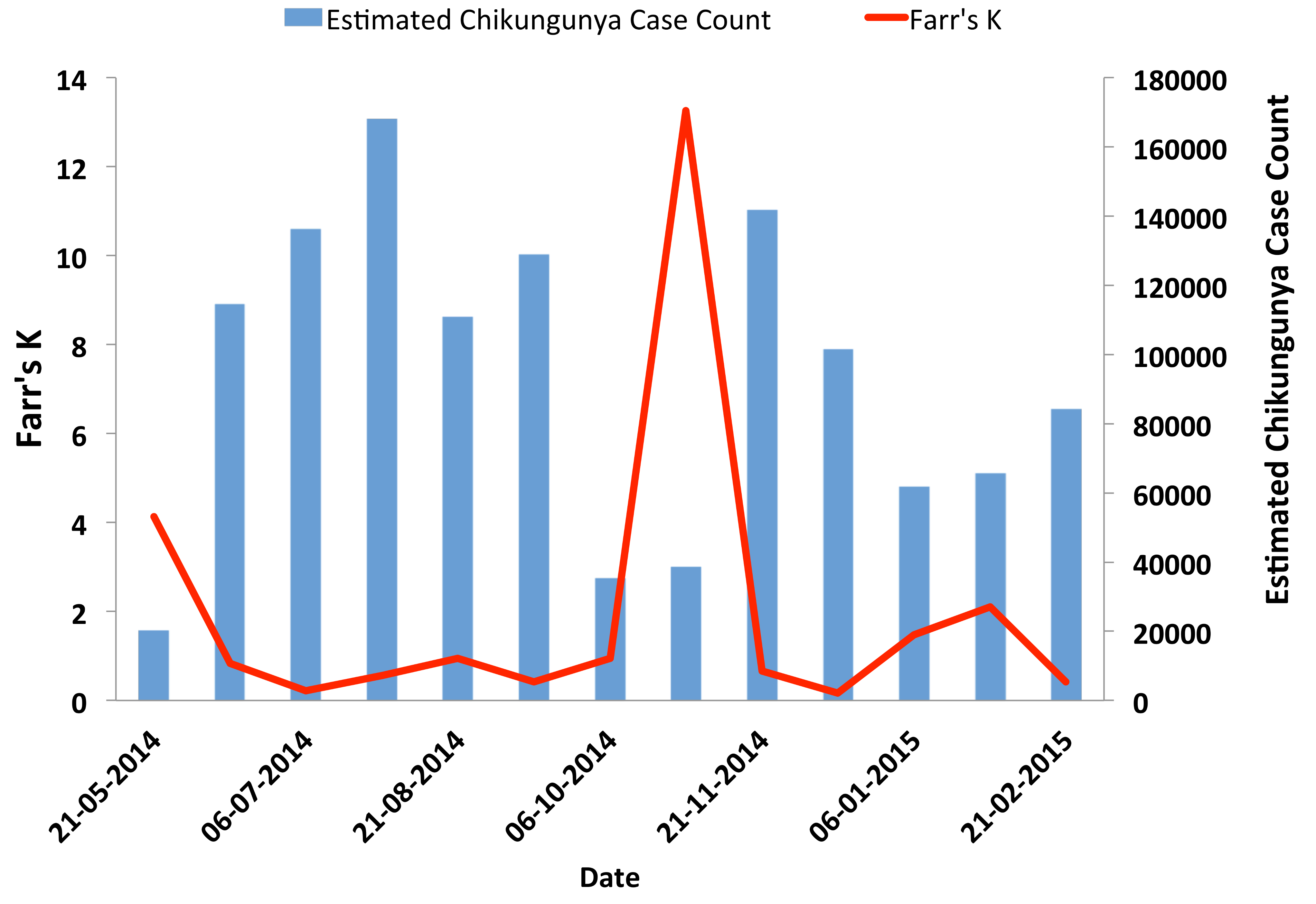}
\caption{\label{fig:Figure 7} A possible application for raw estimates of Farr's $K$ emerged in analysis of data from the 2014-2015 Western Hemisphere Chikungunya outbreak; here it appears that a multi-wave epidemic is signaled by a sudden surge in $K$ to a value $> 1$ (red line), indicating that there is renewed exponential growth in cases (blue bars), rather than exponential decline. X-axis, date of most recent generation; left Y-axis, Farr's $K$; right Y-axis, estimated per-generation Chikungunya case count and transforming resultant estimates.  As in Figure 1, volatile estimates of $K$ were derived using 4-generation series (green diamonds), but estimates of $K$ derived as geometric means of all available $K$ values (red squares) provided a reasonable approximation of $d$.  }
\label{fig:Figure7}
\end{figure}

\end{document}